\documentclass[aps,prl,reprint,superscriptaddress,nofootinbib]{revtex4-2}

\usepackage{amsmath,amssymb,bm}
\usepackage{graphicx}
\usepackage{dcolumn}
\usepackage{appendix}
\usepackage{tikz}
\usepackage{pgfplots}
\usepackage{xcolor}
\usepackage{epsf}
\usepackage{epsfig}
\usepackage{graphicx}
\usepackage{dcolumn}
\usepackage{braket}
\usepackage{bm}
\usepackage{amsfonts}
\usepackage{amsmath}
\usepackage{amssymb}
\usepackage{color,soul}
\usepackage{wasysym}
\usepackage{mathrsfs}
\usepackage{natbib}
\usepackage{float}
\usepackage{multirow}
\usepackage{mathtools}
\usetikzlibrary{spy}

\begin{document}


\title{Exceptional-Point Geometry of Weak Topological Boundary States}

\author{Rafael A. Molina}
\affiliation{Instituto de Estructura de la Materia (IEM-CSIC), Madrid, Spain}

\author{Kevin A. Gonz\'alez}
\affiliation{Departamento de F\'\i sica, Universidad T\'ecnica Federico Santa Mar\'\i a, Casilla 110 V, Valpara\'\i so, Chile}
\affiliation{Instituto de F\'\i sica, Pontiﬁcia Universidad Cat\'olica de Valpara\'\i so, Av. Brasil 2950, Valpara\'\i so, Chile}

\author{Pedro A. Orellana}
\affiliation{Departamento de F\'\i sica, Universidad T\'ecnica Federico Santa Mar\'\i a, Casilla 110 V, Valpara\'\i so, Chile}

\begin{abstract}
In this article, we demonstrate that weak topology can be formulated geometrically in terms of exceptional singularities of an analytically continued Bloch Hamiltonian. A general plaquette chiral model in two dimensions serves as a minimal realization of dual weak topology, possessing two independent families of weak topological invariants, one for each spatial direction. The weak-topological edge states correspond to exceptional points in complex momentum space, while corner zero modes emerge from exceptional curves obtained by complexifying both momenta. Compact localized states arise when the exceptional roots collapse to the origin. This framework provides a unified complex-momentum description of edge, corner, and compact localization.
\end{abstract}

\maketitle

Topological band theory establishes a connection between robust boundary states and bulk invariants defined in momentum space~\cite{HasanKane2010,QiZhang2011}. In one dimension, the Su--Schrieffer--Heeger (SSH) model serves as the canonical chiral example, where the winding of a complex Bloch function predicts zero-energy end modes~\cite{Su1979,Ryu2002}. In higher dimensions, weak topology emerges when a system can be decomposed into families of lower-dimensional topological subsystems~\cite{Fu2007,Moore2007,Ringel2012}. The corresponding invariants are momentum-resolved, and the associated boundary modes may appear only within finite momentum intervals, depending on the boundary orientation~\cite{Ahn2019,Ghosh2021,Jeon2022,Agrawal2023}. Weak topological phases, therefore, occupy an intermediate position between conventional topological insulators and higher-order topological phases. Their boundary states are protected by lower-dimensional topological structures and depend explicitly on the sample orientation. Although the associated bulk invariants are well understood, a unified geometric description of edge, corner, and compact localized states is still lacking. Existing formulations rely on momentum-resolved winding numbers defined separately for each ribbon geometry, whereas edge, corner, and compact localized states are usually discussed using different theoretical languages. It is therefore unclear whether these localization phenomena admit a common geometric description. Developing such a description would clarify the relationship between weak topology, localization, and the increasing significance of non-Hermitian concepts in topological band theory \cite{Yao2018,BergholtzBudichKunst2021,OkumaSato2023}.

Complex momentum offers an alternative framework for describing boundary localization. In systems with open boundaries, the momentum component perpendicular to the boundary is replaced by a complex variable that characterizes evanescent decay into the bulk. Transfer-matrix methods exploit this structure to identify physical boundary modes~\cite{MongShivamoggi2011}. More recently, analytically continued Bloch Hamiltonians have been demonstrated to exhibit exceptional points and branch singularities, which encode boundary states in Hermitian systems and establish connections to non-Hermitian band theory~\cite{Gonzalez2016,Gonzalez2017,Molina2018,Yao2018,Yokomizo2019,BergholtzBudichKunst2021,OkumaSato2023}.
In this article, we demonstrate that weak bulk-boundary correspondence can be described through a simple geometric formulation involving exceptional singularities of an analytically continued Bloch Hamiltonian. In this framework, edge states, corner states, and compact localized states represent distinct manifestations of a unified complex-momentum structure.

\begin{figure}[t]
\includegraphics[width=\columnwidth]{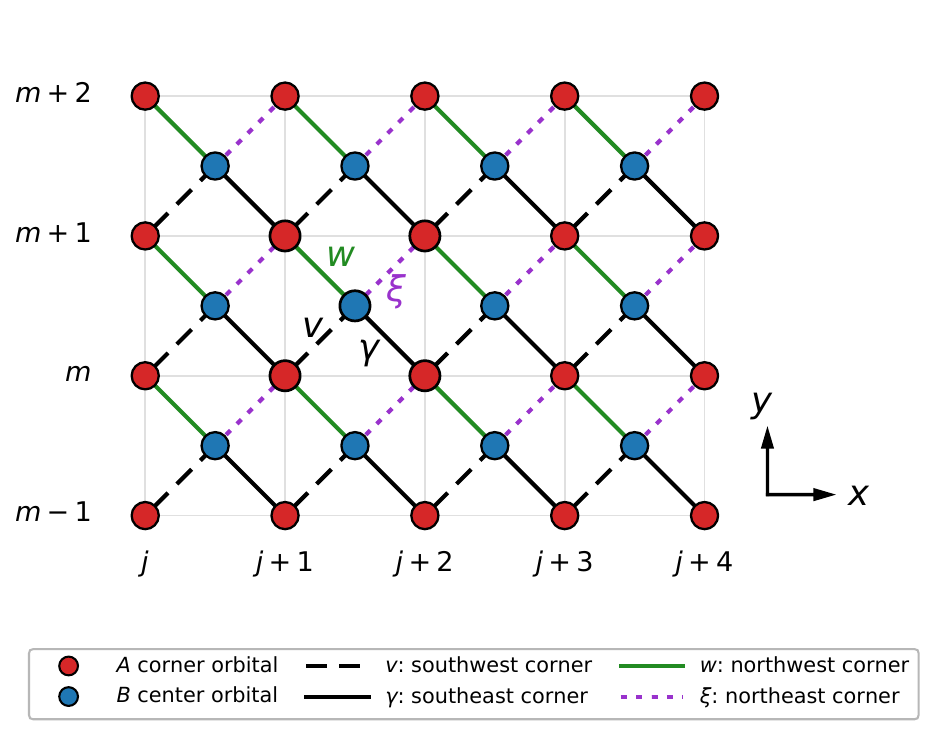}
\caption{\label{fig:PCL_scheme} Lattice schematic of the PCL model with the different hoppings. Sublattice A, red dots, only connects with sites in sublattice B, blue dots. }
\end{figure}


We introduce a minimal Plaquette Chiral Lattice (PCL) model, which is a two-dimensional chiral lattice comprising two sublattices per unit cell. This model features one plaquette-center orbital (B) and one corner orbital (A), with four independent center-to-corner hopping parameters $v$, $\gamma$, $w$, and $\xi$, as illustrated in Fig. \ref{fig:PCL_scheme}. The lattice Hamiltonian is expressed as follows:
\begin{eqnarray}
H_{\mathrm{PCL}}
=
\sum_{m,j}^{W}
(v\,  b^{\dagger}_{j,m} a_{j,m}
+
\gamma\, b^\dagger_{j,m} a_{j+1,m}
 \nonumber \\
+w b^\dagger_{j,m} a_{j,m+1} 
+\xi b^\dagger_{j,m} a_{j+1,m+1}
+\mathrm{h.c.}).
\label{eq:HPCL}
\end{eqnarray}
With periodic boundary conditions, the Bloch Hamiltonian takes the off-diagonal form
\begin{eqnarray}
 h({\bf k})&= &
 \begin{pmatrix}
 0&q({\bf k})\\
 q^*({\bf k})&0
 \end{pmatrix}, \\
 q&= &v+\gamma e^{-ik_x}+w e^{-ik_y}+\xi e^{-i(k_x+k_y)} .
\label{eq:qbulk}
\end{eqnarray}

\begin{figure}[t]
\centering
\includegraphics[width=\columnwidth]{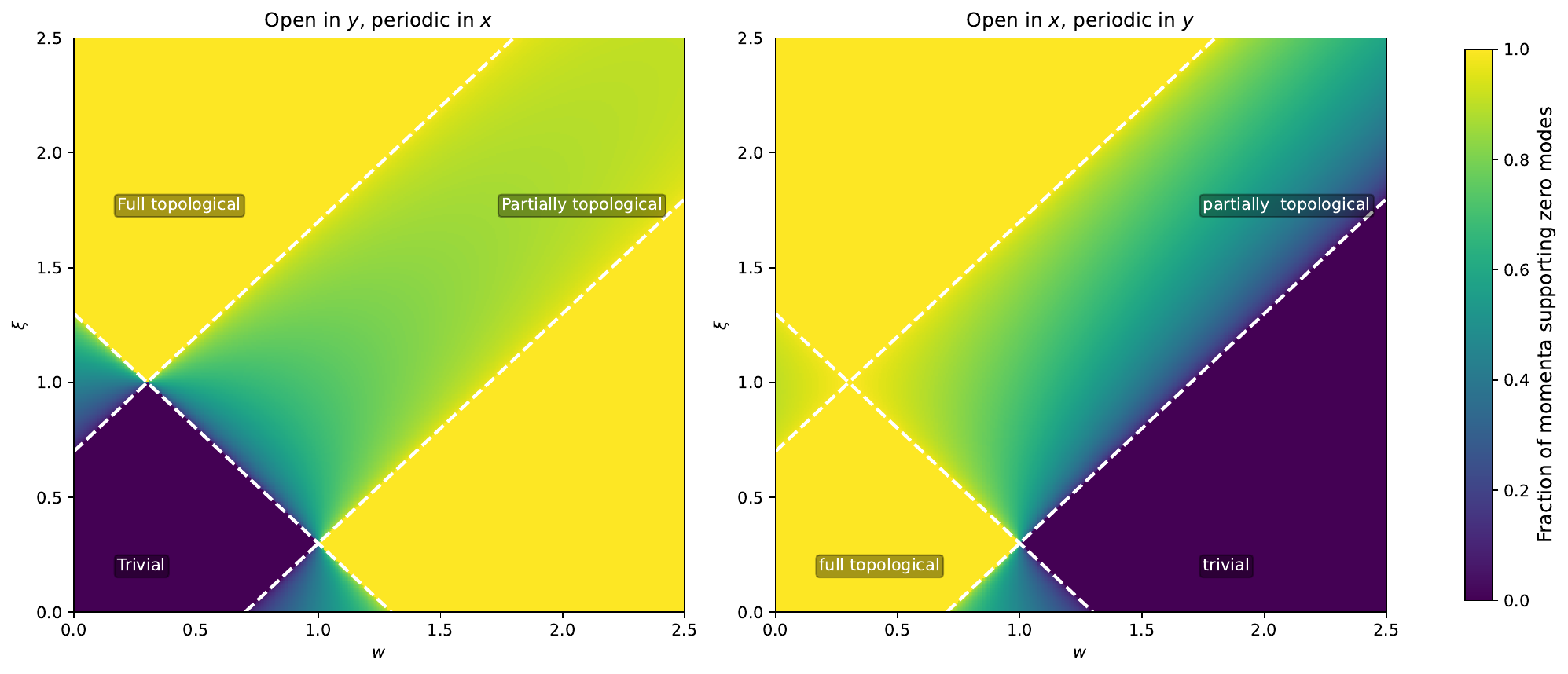}
\caption{Dual weak topology in the PCL model. The color scale gives the fraction of conserved momenta satisfying the SSH condition for ribbons open in $y$ (left) and open in $x$ (right). Dark regions are trivial, bright regions are fully weak topological, and intermediate colors correspond to partial weak topology with boundary states only in finite momentum windows.}
\label{fig:phase}
\end{figure}

The model is a minimal realization of dual weak topology. It also constitutes a natural two-dimensiona genaralization of the SSH chain. It preserves the defining ingredients of the SSH model -chiral symmtery, two sublattices per unit cell, nearest-neighbor inter-sublattice hopping and an off-diagonal Bloch Hamiltonian determined by a single Laurent polynomial- while extending the latter from one to two complex variables. As a consequence, teh familiar one-dimensional winding description acquires two independent weak-topological decompositions associated with two lattice directions,
\begin{equation}
 q=A_x(k_x)+B_x(k_x)e^{-ik_y}
     =A_y(k_y)+B_y(k_y)e^{-ik_x},
\label{eq:decompositions}
\end{equation}
where
\begin{equation}
 A_x=v+\gamma e^{-ik_x},\quad B_x=w+\xi e^{-ik_x},
\end{equation}
with $A_y=v+w e^{-ik_y}$ and $B_y=\gamma+\xi e^{-ik_y}$ obtained by exchanging $x$ and $y$. For a ribbon periodic in $x$ and open in $y$, the momentum-resolved winding number is
\begin{equation}
 \nu_x(k_x)=
 \begin{cases}
 1, & |A_x(k_x)|<|B_x(k_x)|,\\
 0, & |A_x(k_x)|>|B_x(k_x)|.
 \end{cases}
\label{eq:winding}
\end{equation}
The dual invariant $\nu_y(k_y)$ follows analogously. Thus the system can be trivial, weak topological in one orientation only, or dual weak topological when the two families coexist. Weak topology along x arises when the east couplings dominate over the west couplings, weak topology along y arises when the north couplings dominate over the south couplings,
dual weak topology appears when both occur simultaneously, corner states arise from the simultaneous localization in both directions. Figure~\ref{fig:phase} summarizes this structure through the fraction of conserved momenta supporting zero modes for the two ribbon orientations. 

\begin{figure*}[t]
\centering
\includegraphics[width=0.3\textwidth]{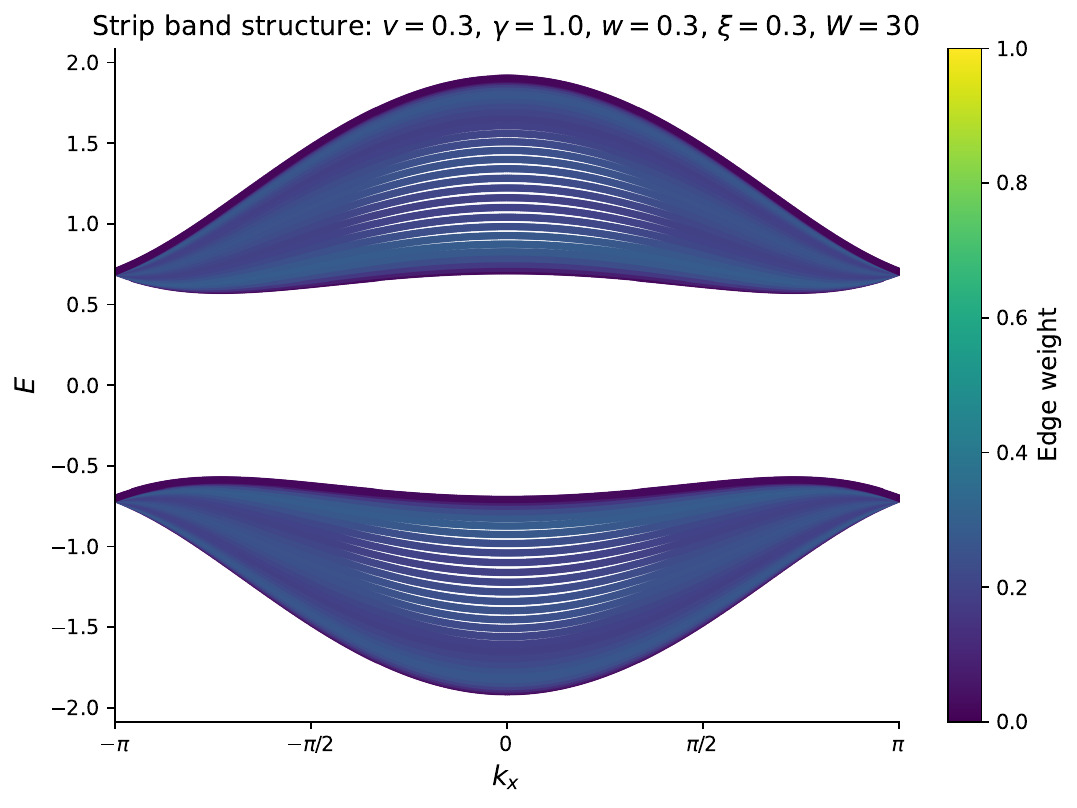}
\includegraphics[width=0.3\textwidth]{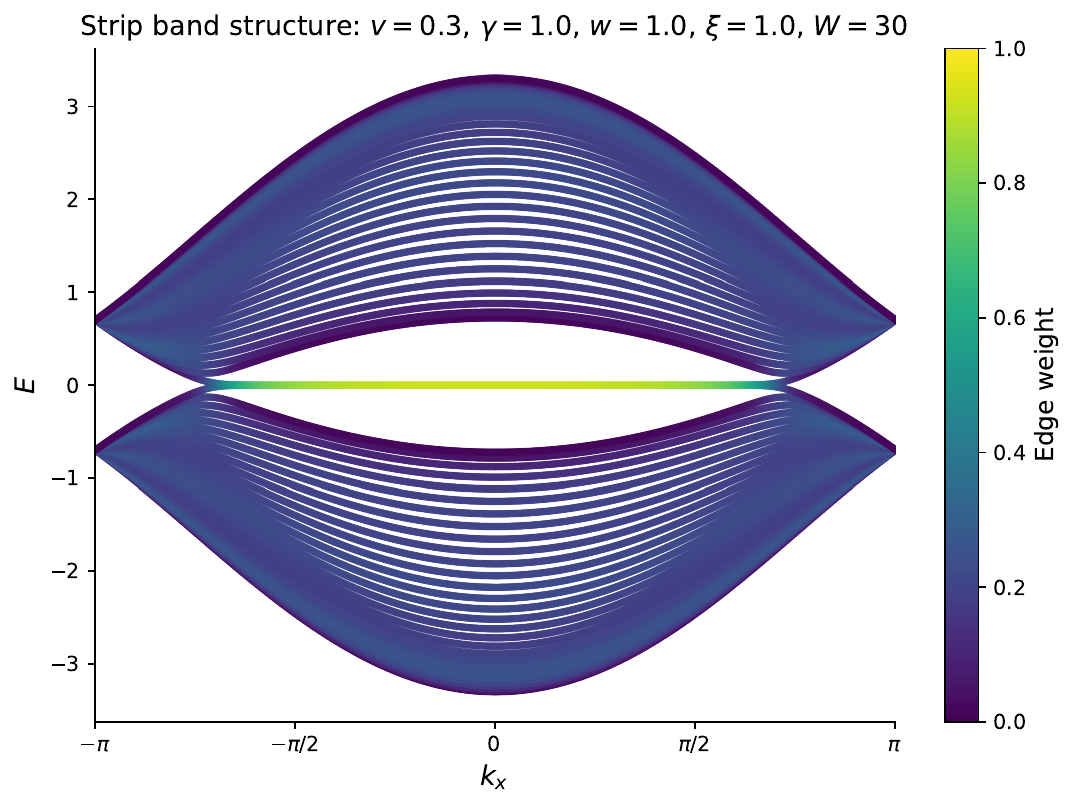}
\includegraphics[width=0.3\textwidth]{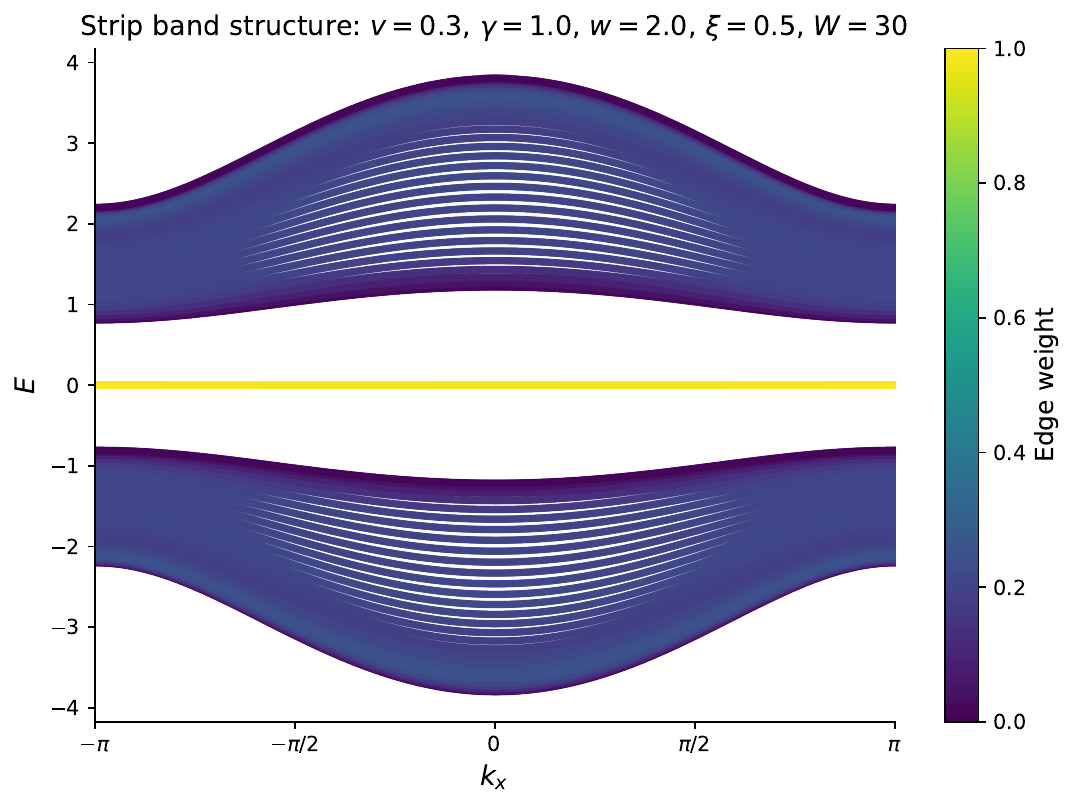}
\caption{\label{fig:strip}
Strip band structures in different topological regimes computed numerically from the finite lattice Hamiltonian periodic in $x$ and open in $y$ with $W$ number of sites. Edge states appear within momentum windows determined by the SSH conditions. We only show examples with a finite y direction, equivalent spectra can be found in the dual direction exchanging the parameters $w$ and $\gamma$. The color intensity represents the weight of the wave function in the three sites closer to the edges, directly related to the localization length. }
\end{figure*}

In Fig. \ref{fig:strip} we show examples of the band structure of one strip on the different phases, from left to right, trivial, mixed and weak topological. 

The exceptional-point formulation follows by complexifying the transverse momentum. For a $y$-open ribbon we define $z_y=e^{-i(k_y+i\alpha_y)}$, so that
\begin{equation}
 q_+(k_x,z_y)=A_x(k_x)+B_x(k_x)z_y .
\end{equation}
The analytically continued Hamiltonian is non-Hermitian because the reciprocal block is no longer the complex conjugate of $q_+$. Zero-energy solutions satisfy
\begin{equation}
 q_+(k_x,z_y)=0,
 \qquad
 z_y^{\rm EP}(k_x)=-\frac{A_x(k_x)}{B_x(k_x)} .
\label{eq:zep}
\end{equation}
At these points $q_+=0$ while the reciprocal block is generically nonzero; the zero-energy degeneracy is defective and therefore exceptional. The physical boundary condition selects the roots inside the unit disk,
\begin{equation}
 |z_y^{\rm EP}(k_x)|<1,
\end{equation}
which is precisely the weak-topological condition in Eq.~\eqref{eq:winding}. The reciprocal root, related by $z_y\rightarrow1/z_y^*$ for real hoppings, describes the opposite edge. At the transition $|z_y^{\rm EP}|=1$, the reciprocal exceptional points meet on the real Brillouin zone, the degeneracy becomes Hermitian, the bulk gap closes, and the localization length diverges.

\begin{figure}[t]
\centering
\includegraphics[width=0.47\columnwidth]{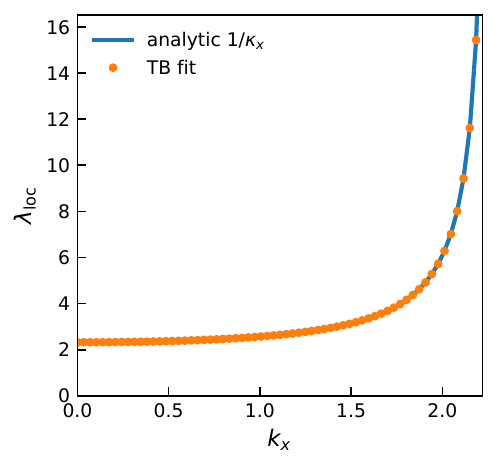}
\includegraphics[width=0.47\columnwidth]{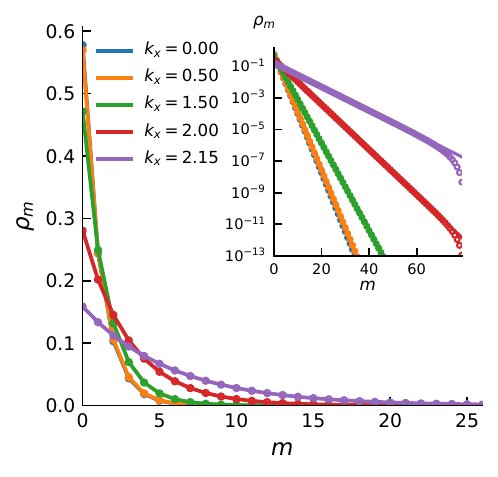}
\caption{Left: analytic localization length $\lambda_x=(\log|B_x/A_x|)^{-1}$ and compared with tight-binding fits. The parameters are $v=0.3$, $\gamma=1.0$, $w=1.0$, $\xi=1.0$, and $W=80$ Right: edge-state density profiles for representative momenta. Solid curves show $\rho_m\propto |z_y^{\rm EP}|^{2m}$ and markers are numerical tight-binding densities; the logarithmic inset displays the exponential decay.}
\label{fig:edge}
\end{figure}

The same root gives the full spatial profile of the boundary state. Since $\psi_m\sim(z_y^{\rm EP})^m$, the decay rate is
\begin{equation}
 \kappa_x(k_x)=-\log|z_y^{\rm EP}(k_x)|
 =\log\left|\frac{B_x(k_x)}{A_x(k_x)}\right|,
\qquad
 \lambda_{\rm loc}=\kappa_x^{-1} .
\label{eq:kappa}
\end{equation}
Figure~\ref{fig:edge} compares this analytic prediction with tight-binding eigenstates of finite ribbons. The agreement of both the fitted localization length and the full density profile confirms that the exceptional root is not merely a formal construction: it quantitatively determines the physical weak-topological boundary mode.

\begin{figure*}[t]
\centering
\includegraphics[width=0.95\textwidth]{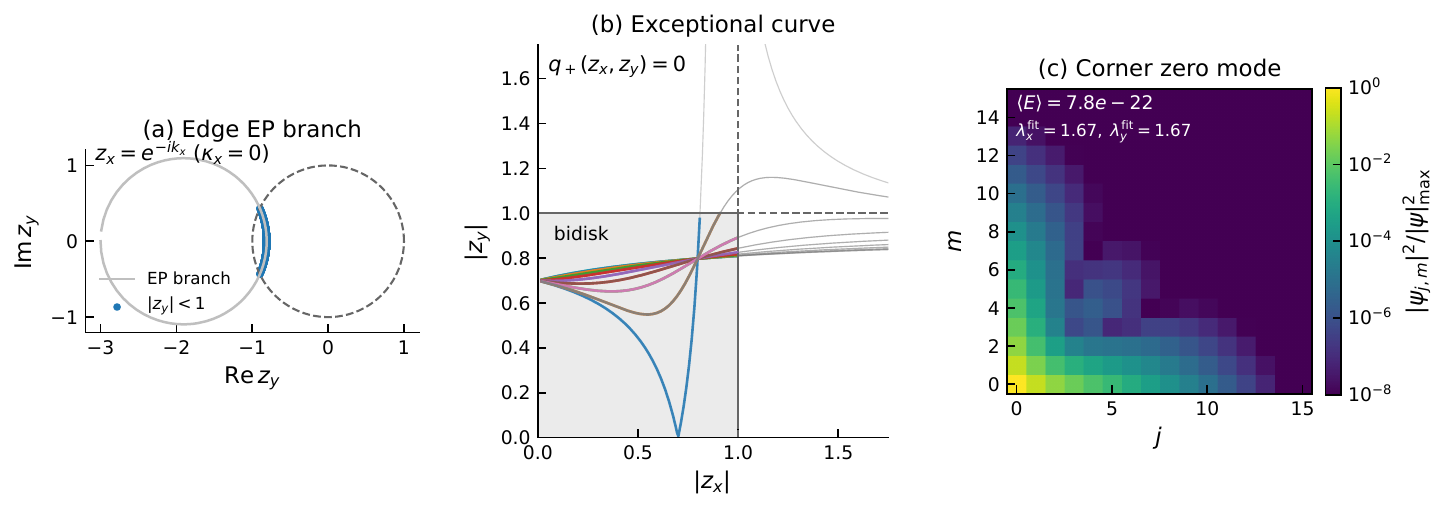}
\caption{Exceptional curves and corner-localized zero modes. (a) Edge exceptional-point branch obtained from $z_y=-(v+\gamma z_x)/(w+\xi z_x)$ with $z_x=e^{-ik_x}$. The highlighted part satisfies $|z_y|<1$. (b) Projection of the two-complex-momentum exceptional curve $q_+(z_x,z_y)=0$ onto the $(|z_x|,|z_y|)$ plane. The shaded square is the bidisk where solutions decay in both directions. (c) Density of a near-zero eigenstate of a finite open PCL sample, localized near a corner.}
\label{fig:curve}
\end{figure*}

For a finite sample open in both directions, neither momentum is conserved. The natural extension is to complexify both crystal momenta,
\begin{equation}
 z_x=e^{-i(k_x+i\alpha_x)},\qquad z_y=e^{-i(k_y+i\alpha_y)} .
\end{equation}
The zero-energy condition becomes
\begin{equation}
 q_+(z_x,z_y)=v+\gamma z_x+w z_y+\xi z_xz_y=0 .
\label{eq:curve}
\end{equation}
Unlike the strip problem, Eq.~\eqref{eq:curve} defines a complex one-dimensional manifold,
\begin{equation}
 {\cal E}_+=\{(z_x,z_y)\in\mathbb C^2:q_+(z_x,z_y)=0\},
\end{equation}
which is an exceptional curve of the analytically continued chiral Hamiltonian whenever the reciprocal block is nonzero. Solving Eq.~\eqref{eq:curve} gives
\begin{equation}
 z_y=-\frac{v+\gamma z_x}{w+\xi z_x},
 \qquad
 z_x=-\frac{v+wz_y}{\gamma+\xi z_y} .
\end{equation}
The strip result is recovered by constraining one variable to the unit circle. A corner state, however, must decay in both directions. For a lower-left corner, the asymptotic form $\psi_{j,m}\sim z_x^jz_y^m$ requires
\begin{equation}
 |z_x|<1,\qquad |z_y|<1 .
\end{equation}
Thus corner-localized zero modes are selected by the portion of the exceptional curve lying inside the bidisk. In compact form,
\begin{equation}
 \text{edge state}\Longleftrightarrow {\cal E}_+\cap\{|z_y|<1, |z_x|=1\},
\end{equation}
whereas
\begin{equation}
 \text{corner state}\Longleftrightarrow {\cal E}_+\cap\{|z_x|<1, |z_y|<1\} .
\end{equation}
The reciprocal curve $\mathcal E_-$ describes the opposite corner. Figure~\ref{fig:curve} illustrates this construction and compares it with a finite-sample zero mode. Although corner modes are often discussed in the language of higher-order topology~\cite{Benalcazar2017,Schindler2018,Li2022}, here they arise from the simultaneous weak-topological structure of the two ribbon orientations rather than from a quantized quadrupole moment. The conventional weak bulk-boundary correspondence is therefore equivalent to selecting the physically admissible portion of the exceptional manifold in complex momentum space.

\begin{figure}[t]
\centering
\includegraphics[width=0.9\columnwidth]{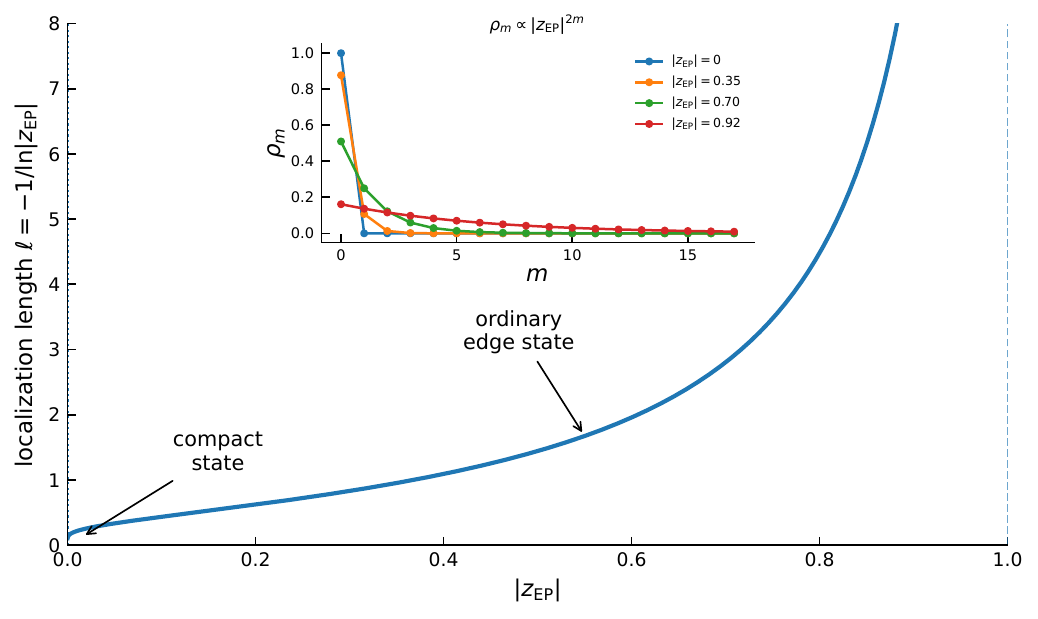}
\caption{Compact localization as a singular limit of the exceptional root. The localization length is $\ell=-1/\ln |z_{\rm EP}|$. The origin gives a compact state, $0<|z_{\rm EP}|<1$ gives exponential localization, and $|z_{\rm EP}|\to1$ gives delocalization. The inset shows representative densities $\rho_m\propto |z_{\rm EP}|^{2m}$.}
\label{fig:compact}
\end{figure}

The same geometry also clarifies compact localization. 
For fixed $k_x$ the effective one-dimensional spectrum of the PCL model
\begin{equation}
 E_\pm(k_y)=\pm\sqrt{|A_x|^2+|B_x|^2+2|A_x||B_x|\cos(k_y+\phi)}
\end{equation}
collapses whenever $A_x(k_x)=0$ or $B_x(k_x)=0$. These two flat sectors are topologically distinct for a $y$-open boundary. If $A_x=0$, then $z_y^{\rm EP}=0$ and the edge state is compactly localized at the boundary. If $B_x=0$, then $z_y^{\rm EP}=\infty$, the trivial dimerized limit for that boundary. Thus the compact state is the zero-localization-length limit of the same weak-topological exceptional root,
\begin{eqnarray}
 z_y^{\rm EP}=0\Rightarrow\text{compact},\\
 0<|z_y^{\rm EP}|<1\Rightarrow\text{exponential},\\
 |z_y^{\rm EP}|=1\Rightarrow\text{critical} .
\end{eqnarray}
This hierarchy is displayed in Fig.~\ref{fig:compact}. Related Creutz--SSH ladders provide quasi-one-dimensional settings where caging, topology, and zero modes coexist~\cite{Zurita2021}; here the compact limit is embedded in a two-dimensional weak-topological geometry.

We have shown that dual weak topology in the PCL model possesses a geometric structure in complex momentum space. Ordinary edge states correspond to exceptional points inside the unit disk, corner states to exceptional curves inside the bidisk, and compact localized states to roots at the origin. The topological transition occurs when reciprocal exceptional singularities reach the real Brillouin zone, where the localization length diverges, and the boundary state is lost. This provides a unified complex-momentum formulation of weak bulk--boundary correspondence and suggests that exceptional singularities are not only a feature of intrinsically non-Hermitian spectra, but also a natural language for boundary localization in Hermitian weak topological phases. 

More generally, our results suggest that exceptional singularities provide a natural geometric language for weak topology. The exceptional-point description not only reproduces the conventional bulk-boundary correspondence but also organizes different localization phenomena into a single hierarchy. Exponentially localized edge states correspond to exceptional points inside the unit disk, corner states correspond to exceptional curves inside the bidisk, and compact localized states emerge when the corresponding roots collapse to the origin. From this perspective, weak topology can be viewed as a geometric selection principle acting on the singularity structure of the complexified Hamiltonian.

In non-reciprocal extensions, where the analytically continued Hamiltonian becomes the physical non-Hermitian Hamiltonian, this structure should connect directly to non-Bloch band theory and generalized Brillouin zones~\cite{Yao2018,Yokomizo2019}. This connection raises the intriguing possibility that weak topological invariants and non-Bloch topological invariants may ultimately admit a common formulation in complex momentum space.

Our results also suggest several experimental directions. The plaquette chiral lattice introduced here should be viewed as a minimal theoretical realization of the underlying geometry rather than a material-specific model. Similar weak-topological structures naturally arise in photonic waveguide arrays \cite{Rechtsman2013,Mittal2019,Caceres-Aravena2024}, electric-circuit networks \cite{Lee2018,Imhof2018}, acoustic and mechanical metamaterials \cite{Susstrunk2015}, and ultracold atoms in optical lattices \cite{Cooper2019}, where hopping amplitudes and boundary conditions can be engineered with high precision. In these platforms, the predicted exceptional-point geometry could be explored through measurements of boundary localization lengths, momentum-resolved edge spectra, or corner-state wave functions, providing direct experimental access to the complex-momentum formulation of weak bulk-boundary correspondence.

Because the exceptional-point formulation relies only on the analytically continued Bloch Hamiltonian, it should be applicable well beyond the specific PCL model studied here, providing a general geometric framework for weak-topological boundary localization in chiral systems.

\begin{acknowledgments}
This work has been supported by the Agencia Estatal de Investigaci\'on from Spain (Grant PID2022-136285NB-C31). P.A.O. acknowledges support from DGIIE USM PI-LIR-24-10, FONDECYT Grant No. 1230933.
\end{acknowledgments}

\bibliography{References_DWT}

\end{document}